\documentclass[twocolumn,showpacs,preprintnumbers,amsmath,amssymb,nofootinbib,longbibliography]{revtex4-1}

\usepackage{graphicx}
\usepackage{dcolumn}
\usepackage{bm}
\usepackage{liecolon}
\usepackage{units}
\usepackage{mathrsfs}

\begin{document}

\preprint{}

\title{A Hamiltonian Perturbation Theory for the Nonlinear Vlasov Equation}

\author{Stephen D. Webb}%
 \email{swebb@radiasoft.net}
 \homepage{www.radiasoft.net}
\affiliation{RadiaSoft, LLC, 1348 Redwood Ave., Boulder, CO 80304}
 \altaffiliation[]{}

\date{\today}

\begin{abstract}
The nonlinear Vlasov equation contains the full nonlinear dynamics and collective effects of a given Hamiltonian system. The linearized approximation is not valid for a variety of interesting systems, nor is it simple to extend to higher order. It is also well-known that the linearized approximation to the Vlasov equation is invalid for long times, due to its inability to correctly capture fine phase space structures.  We derive a perturbation theory for the Vlasov equation based on the underlying Hamiltonian structure of the phase space evolution. We obtain an explicit perturbation series for a dressed Hamiltonian applicable to arbitrary systems whose dynamics can be described by the nonlinear Vlasov equation.
\end{abstract}

\maketitle

\section{Introduction}

The Vlasov equation~\cite{vlasov:68} describes collisionless ensembles of particles moving in their self-fields and external potentials. In general the equation is insoluble, and the canonical approximation is to assume that the phase space density, $\psi(p, q, t) = \psi_0 + \delta \psi$, where $\psi_0$ is some equilibrium and $\delta f$ is a small perturbation\footnote{
This is the standard treatment in almost every plasma physics textbook. Particular examples include~\citet{ichimaru:text}, \citet{kt_86}, \citet{lifshitz_pitaevskii:text}, and \citet{nicholson_text:83}, although there are many, many other good treatments of the subject.}. 

There are a variety of physical systems for which the linearization approximation is invalid. Systems with large charge separations at the unperturbed level, such as laser plasma accelerators~\cite{*[{}][{, and citations therein.}] esarey_schroeder_leemans:09} or plasma wakefield accelerators~\cite{*[{}] [{, and citations therein.}]joshi:02} in the blowout regime, cannot be modeled as a small perturbation to a thermal distribution. Systems whose unperturbed equilibrium would generate self-fields, such as beams in strong-focusing particle accelerators~\cite{kapchinskij_vladimirskij:59, laslett:63,sacherer:68} or astrophysical systems which experience kinetic relaxation that cannot be correctly described by the Vlasov equation~\cite{lynden-bell:67}. Other systems, with time-varying unperturbed Hamiltonians, may not even have a reasonable equilibrium distribution.

Furthermore, the linearization treatment neglects a term proportional to $\partial_p \delta \psi$, which limits the validity of the approximation to short times, even for small perturbations. After a time $\tau_0$, fine structures can appear in phase space which makes the momentum-derivative of $\psi$ quite large. The linearized Vlasov equation approximates this away in an uncontrolled way. This can be understood as the Vlasov equation is not a fluid equation, it is a statement concerning Hamiltonian flows on phase space. This filamentation can explain saturation dynamics such as in free-electron lasers~\cite{gluckstern_etal:93}.

The purpose for this paper is to derive a new approach to computing the solution of the Vlasov equation in terms of Hamiltonian mechanics, in a manner that allows higher-order approximations to be constructed consistently. This approach transforms the self-consistent problem into a single-particle problem using a modified Hamiltonian dressed by the self-fields of the unperturbed orbits.

\section{Limitations of the Linearized Vlasov Equation}

The nonlinear Vlasov equation is given by
\begin{equation}
\frac{\partial \psi}{\partial t} + \dot{z} \cdot \frac{\partial \psi}{\partial z} = 0
\end{equation}
where $z = (q_1, \dots, q_n, p_1, \dots, p_n)$ are the phase space co\"ordinates and $\dot{z}$ satisfies Hamilton's equations of motion
\begin{equation}
\dot{z}_i = J_{i j} \frac{\partial H}{\partial z_j}
\end{equation}
where repeated indices are summed over and $J$ is the antisymmetric $2n \times 2n$ matrix
\begin{equation}
J = \left (
\begin{array}{cc}
0 & I\\
-I & 0
\end{array}
\right ).
\end{equation}
The standard linearization procedure is to break the Hamiltonian into $H = H_0 + H_1(\psi)$, insert $\psi = \psi_0 + \delta \psi$ where $\psi_0 = \psi_0(H_0)$ is a fixed point of the unperturbed system, assume that $H_1(\psi_0) = 0$, and drop terms $\mathcal{O}(\delta \psi^2)$, leaving
\begin{equation}
\frac{\partial \delta \psi}{\partial t} + J_{i j} \frac{\partial H_0}{\partial z_j} \frac{\partial \delta \psi}{\partial z_i} + J_{ij} \frac{\partial }{\partial z_j}H_1(\delta \psi) \frac{\partial \psi_0}{\partial z_i} = 0.
\end{equation}
The term neglected is
\begin{equation}
\mathcal{O}(\delta \psi^2) = J_{ij} \frac{\partial }{\partial z_j}H_1(\delta \psi) \frac{\partial \delta \psi}{\partial z_i}.
\end{equation}
The most familiar example, a non-relativistic free plasma in its own self-fields, takes the form
\begin{equation}
\frac{\partial \delta \psi}{\partial t} + \frac{p}{m} \cdot \frac{\partial \delta \psi}{\partial q} - \left (\frac{\partial}{\partial z} e\varphi(\delta \psi) \right ) \cdot \frac{\partial \psi_0}{\partial p} = 0.
\end{equation}
This problem is then amenable to various methods in linear partial differential equations, which then treats the ensemble of particles as linear waves in phase space. This treatment cannot account for the full complexity of phase space evolution in Hamiltonian systems.

As noted by Villani~\cite{villani:10}, the linearized Vlasov equation is only valid for short times. Villani ascribes this to the fact that we have neglected a term proportional to $\delta \psi \partial_p \delta \psi$ and, while a function may be small, there is no assurance that its derivative will remain small as well. If the phase space develops fine scale structures, the derivative can become quite large.

\begin{figure}
\includegraphics[scale=0.5]{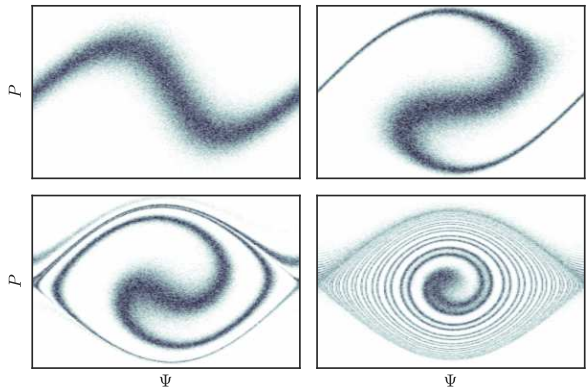}
\caption{Filamentation of trapped particles in a comoving electric field at $t = \nicefrac{1}{\omega_0}$ (upper left), $t=\nicefrac{3}{\omega_0}$ (upper right), $t=\nicefrac{10}{\omega_0}$ (lower left), and $t = \nicefrac{100}{\omega_0}$ (lower right).} \label{filamentation}
\end{figure}

Missing from this analysis is the origin of filamentation: nonlinear Hamiltonian dynamics introduces frequency spread in the single-particle trajectories. Indeed, a perturbing plane wave with an electric field of the form
\begin{equation}
E = E_0 \cos (k x - \omega t)
\end{equation}
in a one-dimensional problem has trapped and untrapped solutions, and the trapped solutions may initially bunch into a sinusoidal charge distribution, but the frequency of revolution in the trapped region varies with amplitude and the distribution eventually filaments. The self-consistent fields will also introduce these nonlinearities, further damping the oscillations and filamenting phase space. If the single-particle trajectories are dominated by a potential with an associated variation in frequency $\Delta \omega$, then within a time $\tau \sim (\Delta \omega)^{-1}$ filamentation will occur and $\partial_p \delta \psi$ will become non-negligible. This is illustrated in the single-particle trajectories of a distribution with a peak in the trapping regime in fig. (\ref{filamentation}). For short times, we see an approximately sine-wave distribution in phase space, but eventually the frequency spreads cause fine structures to form which the linearized Vlasov equation cannot capture.

O'Neil provides an analysis of this problem for an initial traveling electric wave~\cite{oneil:65} and demonstrates this particular problem illustrates Landau damping including the nonlinear dynamics. However, O'Neil never addresses the generic problem -- solving for the characteristics in the nonlinear Vlasov equation -- nor does he provide a systematic approach to higher order approximations. In this paper, we will derive perturbation theory based on the underlying Hamiltonian structure of the dynamics. The resulting perturbation series can then be used to study a self-consistent plasma as single particle dynamics in a dressed potential.

\section{Hamiltonian Mechanics \& Symplectic Maps}

This treatment of the Vlasov equation requires a particular formulation of Hamiltonian mechanics. In this section, we briefly outline the Lie algebraic tools that describe Hamiltonian mechanics. More thorough discussion can be found elsewhere~\cite{dragt_finn:1976,dragt_forest:83,dragt_text:2011,dragt:79,dragt:82}.

Let $f = f(p, q, t)$ be a function of phase space variables. Then each $f$ is associated with a \emph{Lie operator} $\lieop{f}$ whose action on another function $g$ is Poisson brackets:
\begin{equation}
\lieop{f} g = [f, g].
\end{equation}
We define $\{\lieop{f}, \lieop{g}\}$ as the commutator of the Lie operators
\begin{equation}
\{\lieop{f}, \lieop{g}\} = \lieop{f} \lieop{g} - \lieop{g} \lieop{f}.
\end{equation}
This allows us to define the \emph{Lie adjoint} to $\lieop{f}$, $\liead{f}$, which acts on Lie operators by taking commutators:
\begin{equation}
\liead{f} \lieop{g} = \{\lieop{f}, \lieop{g}\}.
\end{equation}
The \emph{Lie transformation} generated by the Lie operator $\lieop{f}$ is defined by the exponential
\begin{equation}
\exp (\lieop{f}) = \sum_{n=0}^\infty \frac{\lieop{f}^n}{n!}
\end{equation}
where $\lieop{f}^n = \lieop{f} (\lieop{f}^{n-1})$ has the action of taking $n$ nested Poisson brackets. 

The action of a Lie operator on a function of phase space is given by
\begin{equation}
\lieop{f} g(z) = g(\lieop{f} z).
\end{equation}
The similarity transformation property of Lie transformations on functions of Lie operators is given by
\begin{equation}
\exp (\lieop{f}) \lieop{g(z)} \exp (-\lieop{f}) = \lieop{g \left ( \exp (\lieop{f}) z\right )}.
\end{equation}

In this formalism, Hamilton's equations state that
\begin{equation}
\dot{z} = - \lieop{H} z
\end{equation}
where $H$ is the Hamiltonian. We define the \emph{symplectic map} $\mathscr{M}$ as the map which takes the initial co\"ordinates $z^i$ to the final co\"ordinates $z$, 
\begin{equation}
z(t) = \mathscr{M}_t z^i.
\end{equation}
Inserting this into Hamilton's equation gives
\begin{equation}
\begin{split}
\frac{d}{dt} \mathscr{M} z^i =& [\mathscr{M} z^i, H(z, t)]\\
=& [\mathscr{M}_t z^i, \mathscr{M} H(z^i, t)] \\
= & \mathscr{M}_t \lieop{-H(z^i, t)} z^i
\end{split}
\end{equation}
which then implies that the map satisfies the differential equation
\begin{equation}
\dot{\mathscr{M}_t} = \mathscr{M}_t \lieop{-H(z^i, t)}.
\end{equation}
All the relevant dynamics are contained in the map.

Now suppose the Hamiltonian can be written as the sum of a dominant term and a small perturbation,
\begin{equation}
H = H_0 + \epsilon H_1.
\end{equation}
Furthermore, suppose the map for $H_0$ is known, so that
\begin{equation} \label{unperturbed}
\dot{\mathscr{M}}^{0} = \mathscr{M}^{0}\lieop{-H_0}.
\end{equation}
We may factor the map into two terms, $\mathscr{M} = \mathscr{M}^{I} \mathscr{M}^0$, where $\mathscr{M}^I$ is the \emph{interaction map} we wish to compute~\cite{dragt_forest:83}. Inserting $\mathscr{M}$ into the differential equation for the map, using the product rule, and invoking eqn. (\ref{unperturbed}) gives
\begin{equation}
\begin{split}
\dot{\mathscr{M}} =& \dot{\mathscr{M}}^I \mathscr{M}^0 + \mathscr{M}^I \dot{\mathscr{M}}^0 \\
= &\dot{\mathscr{M}}^I \mathscr{M}^0 + \mathscr{M}^I \mathscr{M}^0 \lieop{-H_0} \\
=& \mathscr{M}^I \mathscr{M}^0 \lieop{-H_0 - \epsilon H_1}
\end{split}
\end{equation}
which then implies that
\begin{equation}
\begin{split}
\dot{\mathscr{M}}^I =& \mathscr{M}^I \mathscr{M}^0 \lieop{- \epsilon H_1} (\mathscr{M}^0)^{-1}\\
=& \mathscr{M}^I \underbrace{\lieop{-\epsilon H_1 (\mathscr{M}^{0} z^i, t)}}_{\lieop{-H_I}}
\end{split}
\end{equation}
which defines the \emph{interaction Hamiltonian} $H_I$. Our formalism will revolve around computing the interaction map directly and computing from it a modified single-particle Hamiltonian whose solutions are the characteristics for the Vlasov equation with self-fields included.

\section{Maps \& the Vlasov Equation}

Given a map $\mathscr{M}$ for a Hamiltonian $H$, the evolution of any phase space quantity is given by
\begin{equation}
g(z, t) = \mathscr{M}_t g(z^i, 0).
\end{equation}
The Vlasov equation may be stated as
\begin{equation}
\frac{d}{d t} \psi = \frac{\partial \psi}{\partial t} + \dot{z} \cdot \frac{\partial \psi}{\partial z} = 0
\end{equation}
This is a statement that the phase space density $\psi$ is a constant of the motion,
\begin{equation}
\psi(z^i, 0) = \psi(z, t).
\end{equation}
The linearized Vlasov equation violates this conservation law, which is fundamental to the geometric structure of solutions to the Vlasov equation. A perturbation theory which accurately includes the Hamiltonian mechanics of the fundamental problem must derive from this conservation law.

We can phrase this conservation law as
\begin{equation}
\mathscr{M}_t \psi(z^i, 0) = \psi(z, t) = \psi(z^i, 0)
\end{equation}
which then implies that
\begin{equation} \label{vlasoveqn}
\psi((\mathscr{M}_t)^{-1}z^i, 0) = \psi(z, t).
\end{equation}
A Hamiltonian picture of the Vlasov equation should treat the problem as trajectories in phase space. This is the approach used by particle-in-cell computational approaches~\cite{birdsall_langdon:85,hockney_eastwood:89}.

\section{Perturbation Series for a Dressed Hamiltonian}

We now can derive a perturbation series for the effective single-particle Hamiltonian of an ensemble of interacting particles. We assume the exact Hamiltonian is of the form
\begin{equation}
H = H_0 + \epsilon H_1[\psi(z, t)]
\end{equation}
where the brackets indicate that $H_1$ is a functional of the phase space distribution, such as a Green's function over the phase space distribution to compute the collective fields. From the previous section, it is clear that computing the map for this system will contain the full physics with self-interactions while preserving the Hamiltonian structure of the solution.

\subsection{Formulation}

The interaction map is given by
\begin{equation}
\dot{\mathscr{M}}^I = \mathscr{M}^I \left (\mathscr{M}^0 \lieop{-\epsilon H_1[\psi(z, t)]} (\mathscr{M}^0)^{-1} \right )
\end{equation}
For definiteness, we write
\begin{equation}
H_1[\psi(z, t)] = \int dz' dt' \mathcal{G}(z^i, t; z', t') \psi(z', t')
\end{equation}
where $\mathcal{G}$ is the Green's function. Then we are left with the problem
\begin{equation}
\dot{\mathscr{M}}^I = \mathscr{M}^I \lieop{-\epsilon \int dz' dt' \mathcal{G}(\mathscr{M}^{0}_t z^i, t; z', t') \psi(z', t')}
\end{equation}
Similarly, we can insert the Vlasov equation, eqn. (\ref{vlasoveqn}), and note that $\mathscr{M}_t^{-1} = \mathscr{M}_{-t}$, to get the final formulation of the problem
\begin{equation} \label{interactionmap}
\dot{\mathscr{M}}^I = \mathscr{M}^I \lieop{-\epsilon \int dz' dt' \mathcal{G}(\mathscr{M}^0_t z^i, t; z', t') \psi(\mathscr{M}_{-t} z'^i, 0)}.
\end{equation}

The differential equation for the interaction map in eqn. (\ref{interactionmap}) is in form of the nonlinear Magnus problem.

\subsection{The Nonlinear Magnus Problem}

 The conventional Magnus problem~\cite{magnus:54} is the solution of the matrix differential equation
\begin{equation}
\dot{y}(t) = \mathbb{A}(t) y(t)
\end{equation}
by assuming that $y(t) = \exp[\Omega(t)] y(0)$ where $\Omega$ is a matrix. It leads to an iterative series of nesting commutators of $\mathbb{A}(t)$ at different times, with $\Omega$ a series expansion
\begin{equation}
\Omega(t) = \sum_{n=0}^\infty \Omega_n(t)
\end{equation}
 the first two terms being
\begin{subequations}
\begin{equation}
\Omega_1(t) = \int_{0}^t \mathbb{A}(t') dt'
\end{equation}
\begin{equation}
\Omega_2(t) = \int_{0}^t dt' \int_0^{t'} dt'' \{\mathbb{A}(t'), \mathbb{A}(t'') \}.
\end{equation}
\end{subequations}
More can be read in~\cite{blanes:09}. This problem has been applied to Hamiltonian mechanics through the Lie algebraic approach by Oteo and Ros~\cite{oteo_ros:91}.

Our problem explicitly contains the map we are attempting to solve for, so it more closely resembles the nonlinear Magnus expansion derived by Casas and Iserles~\cite{casas_iserles:06}. That problem is the solution of the nonlinear matrix differential equation
\begin{equation}
\dot{y}(t) = \mathbb{A}(t; y(t)) y(t).
\end{equation}
They derived an explicit Magnus expansion type solution to the nonlinear problem, where again the Lie operator techniques used by Oteo and Ros may be applied to the nonlinear Magnus expansion.

\subsection{Deriving a Dressed Hamiltonian}

We define the partial sum of the Magnus exponent as
\begin{equation} \label{magnus_series}
\Omega^{[N]} = \sum_{k = 1}^N \epsilon^k \Omega_k(t)
\end{equation}
such that the $N^{th}$-order approximation to the interaction map takes the form 
\begin{equation}
\mathscr{M}^I_t = \exp \left ( \lieop{\Omega^{[N]}(t)} \right ).
\end{equation}
Using the work by Casas and Iserles, we can compute $\Omega^{[N]}$ explicitly:
\begin{widetext}
\begin{subequations}
\begin{equation} \label{firstorder}
\Omega^{[1]}(t) = - \epsilon \int_0^t d\tau \int dz' dt' \mathcal{G}(\mathscr{M}_{\tau}^0 z^i, t; z', t') \psi(\mathscr{M}_{-t'}^0 z'^i, 0)
\end{equation}
\begin{equation} \label{higherorder}
\Omega^{[N]}(t) = -\epsilon \sum_{n=0}^{N-2} \frac{B_n}{n!} \int_0^t d\tau~ \lieop{\Omega^{[N-1]}(\tau)}^n \int dz'dt' \mathcal{G}(\mathscr{M}_\tau^{0}z, \tau; z', t') \psi(e^{\lieop{\Omega^{[N-1]}(-t')}} \mathscr{M}_{-t'}^{0}  z',0), N \geq 2.
\end{equation}
\end{subequations}
\end{widetext}
where $B_n$ are the Bernoulli numbers. This is not explicitly a power series in $\epsilon$, due to the presence of the $N-1^{th}$ order map in the phase space density.

The formal solution to the map is given by this Magnus expansion
\begin{equation}
\mathscr{M} = \exp \left ( \lieop{\Omega^{[N]}(t)} \right ) \mathscr{M}^0_t.
\end{equation}
For the case when $H_0$ is integrable, this is amenable to a map-based normal form analysis. We can also derive an effective single-particle Hamiltonian from this map, which may be easier to manipulate than the maps themselves.

To do this, we take the time derivative of $\mathscr{M}$ once again:
\begin{equation}
\begin{split}
\dot{\mathscr{M}} =& \left (\frac{d}{dt} \exp \left ( \lieop{\Omega^{[N]}(t)} \right ) \right ) \mathscr{M}^0_t + \\
& \exp \left ( \lieop{\Omega^{[N]}(t)} \right ) \left ( \frac{d}{dt} \mathscr{M}_t^0 \right )\\
=& \exp \left ( \lieop{\Omega^{[N]}(t)} \right )\textrm{iex}\left (\liead{\Omega^{[N]}(t)}  \right ) \frac{\partial}{\partial t} \lieop{\Omega^{[N]}} \mathscr{M}^0_t + \\
&\exp \left ( \lieop{\Omega^{[N]}(t)} \right )  \mathscr{M}_t^0 \lieop{-H_0} \\
=& \mathscr{M} \Lieop{-H_0 + \left (\mathscr{M}_t^{0}\right )^{-1} \left [ \textrm{iex}\left (\lieop{\Omega^{[N]}(t)}  \right ) \frac{\partial\Omega^{[N]} }{\partial t}  \right ] }
\end{split}
\end{equation}
where we have used the fact that $\liead{f} \lieop{g} = \lieop{\lieop{f} g}$, that
\begin{equation}
\frac{d}{d t} e^{\lieop{f(t)}} = e^{\lieop{f(t)}}\textrm{iex}(\liead{f(t)}) \lieop{\dot{f}}
\end{equation}
 and that the exponential integral is defined as
\begin{equation}
\textrm{iex}(x) = \int_0^1 d\sigma ~ e^{\sigma x} = \sum_{k = 0}^{\infty} \frac{x^k}{(k+1)!}.
\end{equation}

We can then truncate this series to order $N$ in $\epsilon$ to obtain a new \emph{dressed Hamiltonian}
\begin{equation}
\mathcal{H} = H_0 - \mathscr{M}_{-t}^0 \textrm{iex} \left (\lieop{\Omega^{[N]}(t)} \right ) \frac{\partial \Omega^{[N]}}{\partial t} + \mathcal{O}(\epsilon^{N+1}).
\end{equation}
Recall eqn. (\ref{magnus_series}) and define
\begin{equation}
\sum_{k = 1}^N \epsilon^k h_k = - \mathscr{M}_{-t}^0\textrm{iex} \left ( \lieop{\Omega^{[N]}(t)} \right ) \frac{\partial}{\partial t} \Omega^{[N]}(t).
\end{equation}
By expansing $\Omega^{[N]}$ in powers of $\epsilon$, and matching powers of $\epsilon$ with $h_k$, we can compute the $N^{th}$ order dressed Hamiltonian
\begin{equation}
\mathcal{H}^{(N)} = H_0 + \sum_{k = 1}^N \epsilon^k h_k.
\end{equation}

Explicitly, to first order
\begin{equation}
\begin{split}
h_1 =& -\mathscr{M}_{-t}^0 \frac{\partial \Omega_1}{\partial t} \\
=& \int dz' dt' \mathcal{G}(z^i, t; z', t') \psi(\mathscr{M}_{-t}^0 z'^i)
\end{split}
\end{equation}
and to second order
\begin{equation}
h_2 = \mathscr{M}_{-t}^0 \frac{\partial \Omega_2}{\partial t} - \frac{1}{2} \lieop{\mathscr{M}_{-t}^0 \Omega_1} \left ( \mathscr{M}_{-t}^0 \frac{\partial \Omega_1}{\partial t} \right ).
\end{equation}
Recall, from eqn. (\ref{higherorder}), that the $N^{th}$ order $\Omega^{[N]}$ is computed along the $(N-1)^{th}$ order trajectories.

The leading order term is easy enough to interpret physically: it is the collective fields generated by the unperturbed dynamics. At second order, we have a more complicated result, which involves the fields generated by the first-order trajectories and various terms representing the interaction of the perturbing fields with themselves.

By truncating the series for $\textrm{iex}(\lieop{\Omega^{[N]}(t)})$ at order $N$, we introduce errors in the Hamiltonian beyond $\mathcal{O}(\epsilon^{N+1})$ where a discrepancy between the dressed Hamiltonian and map dynamics. However, this is a higher order effect than what we can accurately describe. We have thus derived a formal perturbation series for first the transfer map and then a dressed Hamiltonian which can be manipulated to compute the single-particle dynamics.

\section{Discussion \& Applications}

We have derived an explicit perturbation theory for computing a dressed Hamiltonian that includes the self-consistent fields of a collisionless ensemble of particles. The trajectories computed from this Hamiltonian are the characteristics which determines the evolution of the distribution function for arbitrary time. This approach addresses the fundamental limitation of the linearized Vlasov equation as discussed by O'Neil and, later, Villani -- the linearized Vlasov equation as a perturbation on ballistic particle motion cannot account for the nonlinear dynamics of the self-fields.

The formalism has a number of advantages over the linearized Vlasov equation treatment, beyond its applicability for long times. It is straightforward, at least formally, to extend the perturbation theory to arbitrary order. It is not predicated on the existence of an equilibrium distribution -- the initial distribution and the resulting dynamics are decoupled, and each initial distribution introduces a different dressed Hamiltonian. It can also be applied to systems which produce non-perturbative charge separations, so long as the resulting self-consistent fields can be treated as perturbations. It also elucidates the underlying Hamiltonian structure of plasma dynamics, which the linearized Vlasov equation obscures.

We used for our derivation the nonlinear Magnus expansion, which treats the system as a single exponential. This was a matter of convenience. An alternative approach would make use of the Fer expansion~\cite{fer:58}, which is a factored product solution of the form:
\begin{equation}
\mathscr{M} = e^{\epsilon^N \lieop{F_N}} \times \dots \times e^{\epsilon \lieop{F_1}}.
\end{equation}
This approach explicitly factors the problem order by order, which can make explicit at which order a certain physical effect appears. This is not necessarily the case for the Magnus expansion. The Fer expansion has no explicit nonlinear analog to the work by Casas and Iserles. The derivation of a nonlinear explicit Fer expansion would be of great interest for comparing this treatment to the Magnus expansion treatment presented here.

This Hamiltonian approach also introduces a number of questions, beyond the scope of this paper but of interest to various fields. Does there exist an analog to integrability for ensembles of interacting particles?, and furthermore is there a KAM-like theorem on the long-term stability of such distributions? Suppose $H_0$ is periodic, such as in a strong-focusing accelerator lattice. Under what circumstances is the entire map periodic? How robust is this periodicity to variations in the initial distribution? Can this formalism be extended computationally as a novel approach to self-consistent algorithms? We leave these questions to future work.

\section{Acknowledgements}

The author would like to thank Alex Dragt for helpful discussions.

This work was sponsored by the Air Force Office of Scientific Research, Young Investigator Program, under contract no. FA9550-15-C-0031. Distribution Statement A. Approved for public release; distribution is unlimited.

\bibliography{vlasov_perturbation_theory.bib}

\end{document}